\documentclass[journal,twoside,final]{IEEEtran}
\usepackage{amsfonts,amsmath,amssymb,bm}
\usepackage[ruled,vlined,linesnumbered]{algorithm2e}
\usepackage{psfrag,graphicx,epsfig,cite,color}
\usepackage{stfloats}
\usepackage{mathrsfs}

\newtheorem{property}{Property}

\usepackage{mathrsfs}

\def \bmats{\left[\begin{smallmatrix}}
\def \emats{\end{smallmatrix}\right]}\def \bmats{\left[\begin{smallmatrix}}
\def \emats{\end{smallmatrix}\right]}
\def \beqi{\begin{IEEEeqnarray}{rcl}\IEEEyesnumber}
\def \eeqi{\end{IEEEeqnarray}}
\def \inum{\IEEEyessubnumber}

\def \bmat{\begin{bmatrix}}
\def \emat{\end{bmatrix}}
\def \beq  { \begin{equation} }
\def \eeq { \end{equation} }
\def \beqn{ \begin{eqnarray} }
\def \eeqn{ \end{eqnarray} }

\begin{document}


\title{Intelligent Reflecting Surface-Assisted Cognitive Radio System}

\author{Jie Yuan,~Ying-Chang Liang,~\IEEEmembership{Fellow,~IEEE}, Jingon Joung,~\IEEEmembership{Senior Member,~IEEE}, Gang Feng,~\IEEEmembership{Senior Member,~IEEE}, and Erik G. Larsson,~\IEEEmembership{Fellow,~IEEE}
\thanks{This research was supported in part by National Natural Science Foundation of China under Grants 61631005 and U1801261, and in part by the National Research Foundation of Korea (NRF) grants funded by the Korea government (MSIT) (2018R1A4A1023826 \& 2019R1A2C1084168). Part of this work has been submitted to 2020 IEEE International Conference on Communications.}
\thanks{J. Yuan, Y.-C. Liang, and G. Feng are with the National Key Laboratory of Science and Technology on Communications and the Center for Intelligent Networking and Communications (CINC), University of Electronic Science and Technology of China (UESTC), Chengdu 611731, China (e-mail: {jieyuan@std.uestc.edu.cn; liangyc@ieee.org; fenggang@uestc.edu.cn}).}
\thanks{J. Joung is with the School of Electrical and Electronics Engineering, Chung-Ang University, Seoul 06974, South Korea (e-mail: {jgjoung@cau.ac.kr}).}
\thanks{E. G. Larsson is with the Department of Electrical Engineering (ISY), Linköping University, SE-581 83 Linköping, Sweden (email: erik.g.larsson@liu.se).}
}
%

\maketitle

\begin{abstract}
Cognitive radio (CR) is an effective solution to improve the spectral efficiency (SE) of wireless communications by allowing the secondary users (SUs) to share spectrum with primary users. Meanwhile, intelligent reflecting surface (IRS), also known as reconfigurable intelligent surface (RIS), has been recently proposed as a promising approach to enhance energy efficiency (EE) of wireless communication systems through intelligently reconfiguring the channel environment. To improve both SE and EE, in this paper, we introduce multiple IRSs to a downlink multiple-input single-output (MISO) CR system, in which a single SU coexists with a primary network with multiple primary user receivers (PU-RXs). Our design objective is to maximize the achievable rate of SU subject to a total transmit power constraint on the SU transmitter (SU-TX) and interference temperature constraints on the PU-RXs, by jointly optimizing the beamforming at SU-TX and the reflecting coefficients at each IRS. Both perfect and imperfect channel state information (CSI) cases are considered in the optimization. Numerical results demonstrate that the introduction of IRS can significantly improve the achievable rate of SU under both perfect and imperfect CSI cases.
 \end{abstract}

\section{Introduction}\label{sec:intro}

\IEEEPARstart{B}{y} the year 2020, wireless communication is expected to serve approximately $50$ billion devices and each person will be surrounded by more than six devices on average \cite{ericsson2011more}. This will lead to tremendous solicitation of radio resources including bandwidth and energy. Therefore, spectral efficiency (SE) and energy efficiency (EE) are becoming two essential criteria for designing future wireless networks \cite{JoHoSu14JSAC}.

Cognitive radio (CR) has been proposed as an effective technique to enhance the SE \cite{mitola1999cognitive,haykin2005cognitive}. Specifically, a spectrum sharing-based CR network allows secondary users (SUs) to share the spectrum with primary users (PUs) while controlling the interference leakage to PU receivers (PU-RXs). One design strategy is to maximize the achievable rates of SUs, while sustaining the interference temperature (IT) at PU-RXs below a certain threshold \cite{haykin2005cognitive}. The threshold in the IT constraint is evidently designed to ensure that the presence of SUs incurs an acceptable degradation in the quality-of-service (QoS) of the PUs. Under this circumstance, the beamforming technique is generally recognized as an effective means to support the optimal transmission scheme \cite{cumanan2009sinr,yiu2008interference}. 
In practice, the CSI obtained by the SU-TX could be imperfect owing to inevitable practical issues, such as channel estimation errors, quantization errors, and outdated information due to feedback delay. To circumvent this issue, the robust beamforming design has been studied to explicitly take into account those errors \cite{zheng2010robust,gong2013robust,suraweera2010capacity,kang2011optimal,rezki2012ergodic,xiong2015robust}.

On the other hand, intelligent reflection surface (IRS), also known as reconfigurable intelligent surface (RIS), has recently attracted considerable attention from the research community of wireless communications for its ability to improve EE \cite{nature18,cui2014coding, BaReRoDeAlZh19,huang2019reconfigurable,liang2019large}. The IRS is an artificial surface made of electromagnetic material that consists of a large number of passive and low-cost reflecting elements \cite{nature18,cui2014coding,BaReRoDeAlZh19,liang2019large}, which introduce phase shifts and amplitude variations of the incident signals. By doing so, the incident electromagnetic wave can be directed to the desired directions \cite{BaReRoDeAlZh19,huang2019reconfigurable}.
Similar to the cooperative relaying systems \cite{ngo2014multipair,JoCh19}, IRS can construct additional wireless links, yet it does not require active radio frequency components. Besides, owing to the passive structure, there is nearly no additional power consumption and added thermal noise during reflection. On the other hand, compared to backscatter communication \cite{yang2017modulation,XiGuLi19,liang2019large}, the IRS aims to assist the transmission of the intended wireless network without attempting its own information transmission.

%
%


Motivated by the inspiration of CR and IRS, in this paper, we propose an IRS-assisted downlink MISO CR network to enhance both SE and EE. As shown in Fig. \ref{systemmodel}, generalizing from a single IRS scenario \cite{YuLiJoFeLa19}, multiple IRSs are deployed to assist the CR network, in which a SU transmitter (SU-TX) with multiple antennas communicates with a single-antenna SU-RX and shares the same spectrum with several PUs.
Our objective is to jointly optimize the beamforming vector at SU-TX and the reflecting coefficients at each IRS for the IRS-assisted CR system to maximize the achievable rate of SU subject to a total power constraint on SU-TX and IT constraints on PU-RXs. Both perfect CSI and imperfect CSI cases are considered. The main contributions of this paper are summarized as follows:


\begin{figure}[!t]
\centering
    \includegraphics[width=0.99\columnwidth]{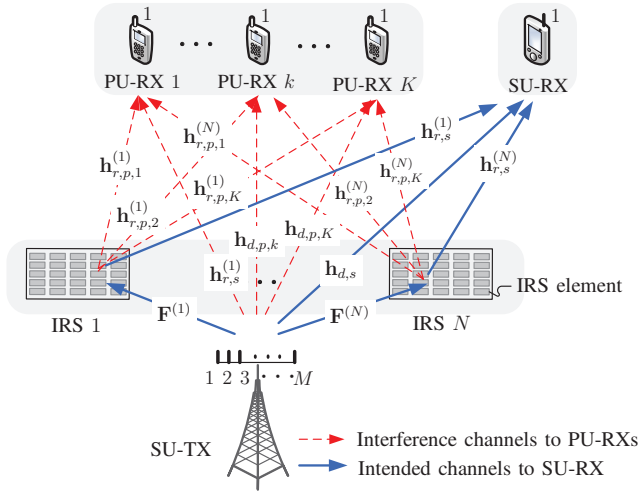}
  \caption{IRS-assisted downlink MISO CR network model with one pair of SU-TX and SU-RS, $N$ IRSs, and $K$ PU-RXs.}\label{systemmodel}
\end{figure}

\begin{itemize}
  \item We formulate a SU rate maximization problem by optimizing the transmit beamforming vector at the SU-TX and reflecting coefficients at each IRS. Both transmit power constraint at SU-TX and interference temperature constraints at PUs are considered.
  \item For perfect CSI case, an efficient iterative algorithm based on the block coordinate descent (BCD) method \cite{wright2015coordinate} is proposed. In each iteration, first, the beamforming vector at the SU-TX is optimized by solving a second order cone programming (SOCP). Then, the reflecting coefficients are optimized by applying the semidefinite relaxation (SDR) technique to relax the rank-$1$ constraint on the reflecting coefficient matrices. Later, the Gaussian randomization scheme proposed in \cite{992139} is used to recover the rank-$1$ variables.
  \item For imperfect CSI case, the worst case approach is used, in which the channel uncertainty is modelled by a given ellipsoid region \cite{zheng2009robust,wang2009worst}. With that, the formulated problem is transformed to the worst-case robust optimization problem, i.e., a maxmin problem. The SDR technique is again used to relax the rank-$1$ constraints of the optimization problem. Then, we transform the maxmin problem to a maximization problem, and the variables are alternatively updated by decoupling the original problem into two subproblems. In each subproblem, using the $\mathcal{S}$-Procedure \cite{beck2006strong}, the equivalent IT and signal-to-noise ratio (SNR) constraints can be derived.

\end{itemize}

 %

The remaining parts of this paper are organized as follows. Section \ref{sec:related work} presents a brief overview of the related work. The system model of IRS-assisted CR system and the problem formulation are introduced in Section \ref{sec:system model}. Section \ref{sec:perfect channel} addresses the perfect CSI case and decouples the original problem into an SOCP subproblem and a semidefinite programming (SDP) subproblem. Section \ref{sec:perfect channel} is devoted to the imperfect CSI case, in which we formulate the worst-case optimization problem and then propose an alternating optimization method to solve it. Numerical results to evaluate the performance of the proposed system are provided and discussed in Section \ref{sec:simulations}. Section \ref{sec:conclusion} concludes this paper.

\textbf{Notations:} The scalars, column vectors, and matrices are denoted by a lower case, boldface lower case, and boldface upper case letters, respectively, i.e., $a$, $\mathbf{a}$, and $\mathbf{A}$. For vector ${\bf a}$, ${\rm diag}({\bf a})$ gives a diagonal matrix whose diagonal elements correspond to ${\bf a}$ and $[{\bf a}]_i$ means the $i$th element of ${\bf a}$. For matrix $\mathbf{A}$, $\mathrm{tr}(\mathbf{A})$, $\|\mathbf{A}\|$, $\mathrm{rank}(\mathbf{A})$, $\mathbf{A}^T$, $\mathbf{A}^\dag$, and $[\mathbf{A}]_{i,j}$ denote its trace, Frobenius norm, rank, transpose, complex conjugate transpose, and the $(i,j)$th element, respectively. $\mathbf{A}\succeq0$ means that $\mathbf{A}$ is a positive semidefinite (PSD) matrix. $\mathbf{A}\otimes\mathbf{B}$ denotes the Kronecker product of $\mathbf{A}$ and $\mathbf{B}$. $\mathrm{vec}(\mathbf{A})$ is a column vector by stacking all the elements of $\mathbf{A}$. $\mathcal{CN}(\mu,\sigma^2)$ denotes the distribution of a circularly symmetric complex Gaussian random variable with mean $\mu$ and variance $\sigma^2$. $\mathbf{0}_{N}$ and $\mathbf{I}_{N}$ denote the $N$-dimensional zero and identity matrices, respectively.


\section{Related Work}
\label{sec:related work}
In this section, we conduct a brief survey for the related work in three aspects, which includes CR beamforming, robust beamforming, and IRS-assisted wireless communication system.
\subsubsection{CR Beamforming}
So far, a lot of CR beamforming techniques have been designed by assuming that the CSI is perfectly known at SU-TX \cite{zhang2008exploiting,cumanan2009sinr,tajer2009beamforming,islam2008joint,yiu2008interference}. Specifically, in \cite{zhang2008exploiting}, a single SU is considered for spectrum sharing by restricting the interference leakages to each PU. For multiple SUs, beamforming techniques are proposed to maximize the minimum signal-to-interference-plus-noise ratio (SINR) and the minimum rate of SUs in \cite{cumanan2009sinr} and \cite{tajer2009beamforming}, respectively.
In \cite{islam2008joint}, centralized joint beamforming and power control methods are studied for multiple SUs with multiple antennas.
The beamforming designed in \cite{yiu2008interference} maximizes the ratio between the received signal power at SUs and the interference leakage to PUs.
\subsubsection{Robust Beamforming}

 Robust beamforming techniques that tackle the channel uncertainty have been addressed in many studies, and the main techniques fall into two categories. One is the stochastic approach, in which the CSI errors are modeled as Gaussian random variables and the parametric constraints are modeled by the probability constraints \cite{zheng2010robust,gong2013robust,suraweera2010capacity,kang2011optimal,rezki2012ergodic,xiong2015robust}. The other one is the worst-case approach, in which the CSI errors are assumed to be within a given uncertainty set and the optimization is performed under the worst-case channel condition \cite{zhang2009robust,zhang2012distributed,cumanan2008robust ,wang2014robust,pei2011secure}.
  Specifically, in \cite{zhang2009robust,zhang2012distributed}, only the CSI uncertainty in primary link is considered for multiple-input single-output (MISO) and multiple-input multiple-output (MIMO) CR network. Considering the CSI uncertainty in both primary and secondary links, the worst-case robust design is used to minimize the total transmit power in \cite{cumanan2008robust}, maximize the robust EE in \cite{wang2014robust}, and maximize the achievable secrecy-rate in \cite{pei2011secure}.

\subsubsection{IRS-assisted Wireless Communication System}
Owing to the promising features of IRS, IRS-assisted wireless communication systems have been vigorously studied recently \cite{yang2019irs,li2019reconfigurable,chen2019intelligent,yu2019enabling,mishra2019channel,
yang2019intelligent,wu2019intelligent,guo2019weighted,NaKaChDeAl19,huang2018energy,
huang2019reconfigurable,han2019large}. Specifically, the achievable rate maximization problem is studied in \cite{yang2019irs} for orthogonal frequency division multiplexing (OFDM) systems. The unmanned aerial vehicle (UAV) trajectory is designed for IRS-assisted UAV system to maximize the average achievable rate \cite{li2019reconfigurable}. The physical layer security problem is considered in \cite{chen2019intelligent} and \cite{yu2019enabling}. The minimum-secrecy-rate maximization problem is studied in \cite{chen2019intelligent} which considers both continuous and discrete reflecting coefficients. The IRS-assisted transmit beamforming is designed in \cite{mishra2019channel} to support wireless power transfer. The IRS-assisted non-orthogonal-multiple access system is considered in \cite{yang2019intelligent}, in which a base station (BS) transmits superimposed downlink signals to multiple users. For multiuser MISO communication systems, the IRS is exploited to enhance the weighted sum-rate in \cite{wu2019intelligent}, \cite{guo2019weighted}, or maximize the minimum SINR in \cite{NaKaChDeAl19}. EE is improved by using transmit power allocation and designing the reflecting coefficients of the IRS aided by precoding \cite{huang2018energy,huang2019reconfigurable}. The impact of phase shift on the ergodic SE is investigated in \cite{han2019large} by exploiting the statistical CSI.

However, there is no related work that introduces multiple IRSs into CR networks while considering both perfect and imperfect CSI cases.

\section{System Model}
\label{sec:system model}

As shown in Fig. \ref{systemmodel}, we consider an IRS-assisted CR downlink network with $N(N\geq1)$ IRSs, $K$ ($K \geq 1$) PU-RXs, and a pair of SU-TX and SU-RX. The SU-TX has $M$ transmit antennas, while each RX has a single receive antenna. We assume that the SU uses the same frequency band as the PUs, and each IRS is composed of $L$ ($L>1$) passive reflecting elements, through which the impinging electromagnetic wave is directed to a desired direction.

\subsection{Channel Model}
We denote the baseband equivalent direct-link channel responses from the SU-TX to PU-RXs and SU-RX by $\mathbf{h}_{d,p,k}\in\mathbb{C}^{M\times 1}$ for $k\in\mathcal{K}=\{1,\ldots,K\}$ and $\mathbf{h}_{d,s}\in\mathbb{C}^{M\times 1}$, respectively, where each element of $\mathbf{h}_{d,p,k}$ and $\mathbf{h}_{d,s}$ is an independent and identically distributed (i.i.d.) complex Gaussian random variable with a zero mean and unit variance, namely, $\mathbf{h}_{d,p,k}\sim\mathcal{CN}({\bf 0}_M,{\bf I}_M)$ and $\mathbf{h}_{d,s}\sim\mathcal{CN}({\bf 0}_M,{\bf I}_M)$, respectively. The baseband equivalent channel responses from the SU-TX to the $n$th IRS is denoted by $\mathbf{F}^{(n)}\in\mathbb{C}^{L\times M}$ for $n\in\mathcal{N}=\{1,\ldots,N\}$. In reality, the IRSs are envisioned to be deployed on the facade of a building close to the BS and users; therefore, we can assume that a line of sight (LoS) path exists in the reflect-link. Therefore, the channel $\mathbf{F}^{(n)}$ can be modeled as a Rician fading channel as follows:
\begin{equation}\label{F}
  \mathbf{F}^{(n)}=\sqrt{\frac{\kappa_1}{\kappa_1+1}}\mathbf{F}^{{(n)},\rm LoS}+\sqrt{\frac{1}{\kappa_1+1}}\mathbf{F}^{{(n)},\rm NLoS},
\end{equation}
where $\kappa_1$ is the Rician factor and $\mathbf{F}^{{(n)},\rm LoS}\in\mathbb{C}^{L\times M}$ represents the fixed channel related to the LoS component, while $\mathbf{F}^{{(n)},\rm NLoS}\in\mathbb{C}^{L\times M}$ represents the NLoS channel matrix, whose elements are i.i.d. that conform to the complex Gaussian distribution with a zero mean and unit variance.

In \eqref{F}, ${\bf F}^{{(n)},\rm LoS}$ is precisely modeled as
\begin{equation}
\mathbf{F}^{{(n)},\rm LoS}=\mathbf{a}_L\left(\beta^{(n)}_\mathrm{AoA}\right)\left(\mathbf{a}_M\left(\beta^{(n)}_\mathrm{AoD}\right)\right)^\dag,
\end{equation}
where $\beta^{(n)}_\mathrm{AoA}$ and $\beta^{(n)}_\mathrm{AoD}$ denote the angle of arrival (AoA) and angle of departure (AoD) of IRS $n$, respectively. Here, $\mathbf{a}_q(\beta)\in\mathbb{C}^{q\times 1}$ is a $q$-dimension uniform linear array response vector given as follows:
\begin{equation}
\mathbf{a}_q(\beta)=\left[1,e^{j\frac{2{\pi}d}{\lambda}\sin(\beta)},e^{j\frac{4{\pi}d}{\lambda}\sin(\beta)},\ldots,e^{j\frac{2(q-1){\pi}d}{\lambda}\sin(\beta)}\right]^T,
\end{equation}
where $\beta$ is the angle, $d$ is the distance between adjacent IRS elements, and $\lambda$ denotes the wavelength of the carrier. To remove the ambiguity in AoA estimation within the AoA interval, we set $d/\lambda$ to be $1/2$.

Likewise, the channels of reflect-link from IRS $n$ to PU-RX $k$ and SU-RX are denoted by $\mathbf{h}^{(n)}_{r,p,k}\in\mathbb{C}^{L\times 1}$ and $\mathbf{h}^{(n)}_{r,s}\in\mathbb{C}^{L\times 1}$, respectively, and they can be modeled as follows:
\beqi
  \mathbf{h}^{(n)}_{r,p,k}&\;=\;&\sqrt{\frac{\kappa_2}{k_2+1}}\mathbf {h}_{r,p,k}^{(n),\rm LoS}+\sqrt{\frac{1}{\kappa_2+1}}\mathbf{h}_{r,p,k}^{(n),\rm NLoS}\!\!\!\!\!\!\!,\\
  \mathbf{h}^{(n)}_{r,s}&=&\sqrt{\frac{\kappa_3}{\kappa_3+1}}\mathbf{h}_{r,s}^{(n),\rm LoS}+\sqrt{\frac{1}{\kappa_3+1}}\mathbf{h}_{r,s}^{(n),\rm NLoS},
\eeqi
where $\kappa_2$ and $\kappa_3$ are the Rician factors of $\mathbf{h}^{(n)}_{r,p,k}$ and $\mathbf{h}^{(n)}_{r,s}$, respectively; $\mathbf {h}_{r,p,k}^{(n),\rm LoS}=\mathbf{a}_M\left(\beta^{(n)}_{r,p,k}\right)\in\mathbb{C}^{L\times 1}$ and $\mathbf{h}_{r,s}^{(n),\rm LoS}=\mathbf{a}_M\left(\beta^{(n)}_{r,s}\right)\in\mathbb{C}^{L\times 1}$ are the fixed LoS channels; $\beta^{(n)}_{r,p,k}$ represents the AoD from the $n$th IRS to the $k$th PU; and $\beta^{(n)}_{r,s}$ represents the AoD from the $n$th IRS to SU. Here, $\mathbf{h}_{r,p,k}^{(n),\rm NLoS}\in\mathcal{C}^{L\times 1}$ and $\mathbf{h}_{r,s}^{(n),\rm NLoS}\in\mathcal{C}^{L\times 1}$ are the NLoS channel vectors, whose elements are i.i.d. and conform to the complex Gaussian distribution.

Before leaving this subsection, we define the primary and secondary compositive channel matrices for SU-RX and PU-RX $k$, respectively, as
\beqi\label{Ch}
\mathbf{H}_s&\;\triangleq\;&\bmat
&\varphi{\rm diag}(\mathbf{h}_{r,s}^{(1),\dag})\mathbf{F}^{(1)} \\
&\varphi{\rm diag}(\mathbf{h}_{r,s}^{(2),\dag})\mathbf{F}^{(2)} \\
&\vdots\\
&\varphi{\rm diag}(\mathbf{h}_{r,s}^{(N),\dag})\mathbf{F}^{(N)} \\
&\varphi\mathbf{h}_{d,s}^\dag\emat\in\mathbb{C}^{(NL+1)\times{M}},\inum\\
\mathbf{H}_{p,k}&\;\triangleq\;& \bmat
&\varphi{\rm diag}(\mathbf{h}_{r,p,k}^{(1),\dag})\mathbf{F}^{(1)} \\
&\varphi{\rm diag}(\mathbf{h}_{r,p,k}^{(2),\dag})\mathbf{F}^{(2)} \\
&\vdots\\
&\varphi{\rm diag}(\mathbf{h}_{r,p,k}^{(N),\dag})\mathbf{F}^{(N)} \\ &\mathbf{h}_{d,p}^\dag\emat\in\mathbb{C}^{(NL+1)\times{M}}.\inum
\eeqi
To avoid clutter in derivation later, we have used $\varphi\ll1$ to denote the relative channel gain between the direct-link and reflect-link channels, which is caused by the “double-fading” effect of the reflect-link channels \cite{guo2019weighted}.

\subsection{Signal Model}
The signals received at the SU-RX and PU-RX $k$, denoted by $y_s$ and ${y}_{p,k}$, respectively, can be written as follows:
\beqi\label{RxSig}
{y}_s&=&\left(\mathbf{h}_{d,s}^\dag+\varphi\sum_{n=1}^{N,\dag}\mathbf{h}^{(n),\dag}_{r,s}\bm{\Phi}^{(n)}{\mathbf{F}^{(n)}}\right)\mathbf{x}_s+{u}_s,
\inum\label{eq:su}\\
{y}_{p,k}&=&\left(\mathbf{h}_{d,p,k}^\dag\!+\varphi\sum_{n=1}^{N}\!\mathbf{h}^{(n),\dag}_{r,p,k}\bm{\Phi}^{(n)}{\mathbf{F}^{(n)}}\right)\mathbf{x}_s\!+\!{u}_{p,k},~~~~~
\inum\label{eq:pu}
\eeqi
where $\bm{\Phi}^{(n)}={\rm diag}(\bm \theta^{(n)})$ and $\bm \theta^{(n)}=\left[\alpha^{(n)}_1 e^{j\theta_1^{(n)}},\ldots,\alpha^{(n)}_L e^{j\theta^{(n)}_L}\right]$ for $n\in\mathcal{N}$ represents the reflecting coefficients of the $n$th IRS. Here, $\theta^{(n)}_l\in[0,2\pi]$ and $\alpha^{(n)}_l\in[0,1]$ denote the phase shifts and amplitude gain induced by the $l$th element of the $n$th IRS, respectively; $\mathbf{x}_s\in\mathbb{C}^{M\times{1}}$ denotes the complex baseband signal transmitted by the SU-TX; ${u}_s\sim\mathcal{CN}(0,\sigma_s^2)$ and ${u}_{p,k}\sim\mathcal{CN}(0,\sigma_{p,k}^2)$ are the additive white Gaussian noise (AWGN) vectors at the SU-RX and PU-RX $k$, respectively; and ${\bf x}_s={\bf w}s$ and $\mathbf{w}\in\mathcal{C}^{M\times{1}}$ represents the fixed transmit beamforming vector for the information symbol $s\sim\mathcal{CN}(0,1)$. By using \eqref{Ch}, the received signals at SU-RX and PU-RX $k$ in \eqref{RxSig} are rewritten as follows:
\beqi\label{RxSig2}
{y}_s&\;=\;&{\bm \theta}^\dag\mathbf{H}_s\mathbf{x}_s+{u}_s,\inum\label{eq:su1}\\
{y}_{p,k}&=&{\bm \theta}^\dag\mathbf{H}_{p,k}\mathbf{x}_s+{u}_{p,k},\inum\label{eq:pu1}
\eeqi
where ${\bm \theta}\triangleq[\bm \theta^{(1)},\bm \theta^{(2)},\ldots,\bm \theta^{(n)},1]^\dag\in\mathbb{C}^{(NL+1)\times{1}}$.

From \eqref{eq:su1}, the instantaneous received SNR at the SU-RX can be rewritten as
\begin{equation}
\begin{split}\label{snr}
\mathsf{SNR}_s=\frac{{\rm E}\left[\left|{\bm \theta}^\dag\mathbf{H}_s\mathbf{x}_s\right|^2\right]}{{\rm E}\left[{u}_s^2\right]}
=\frac{|{\bm \theta}^\dag\mathbf{H}_s\mathbf{w}|^2}{\sigma_{s}^2}.
\end{split}
\end{equation}
Similarly, from \eqref{eq:pu1}, we can derive the interference power at the $k$th PU-RX caused by SU-TX, which is known as IT in CR networks, as
\begin{equation}
\begin{split}\label{it}
\mathsf{IT}_{p,k}={\rm E}\left[\left|{\bm \theta}^\dag\mathbf{H}_{p,k}\mathbf{x}_s\right|^2\right]
=\left|{\bm \theta}^\dag\mathbf{H}_{p,k}\mathbf{w}\right|^2.
\end{split}
\end{equation}

\subsection{Discussion on Channel Estimation}\label{estimation}
Generally, channel estimation is necessary to jointly optimize the reflecting coefficients and the beamforming vector. Considering the passive nature of the IRS, we adopt the time division duplex protocol and use the received uplink pilot signals from the SU-RX and PU-RXs
to estimate the downlink channels, which exploits the channel reciprocity.
 Moreover, following the channel estimation protocol in \cite{mishra2019channel}, the SU-TX only needs to estimate the compositive channel matrices $\mathbf{H}_s$ and $\{\mathbf{H}_{p,k}\}$ instead of estimating $\mathbf{h}_{d,s}$, $\mathbf{h}_{d,p,k}$, $\mathbf{h}_{r,s}^{(n)}$, $\mathbf{h}_{r,p,k}^{(n)}$, and $\mathbf{F}^{(n)}$, respectively.

To estimate the secondary compositive channel matrices $\mathbf{H}_s$, totally $NL+1$ orthogonal pilot signals need to be transmitted by the SU-RX in $NL+1$ time slots. During the first time slot, all elements of each IRS are turned off and the SU-TX estimates the $(N\times L+1)$th column of $\mathbf{H}_s^\dag$, i.e., the direct channel $\mathbf{h}_{d,s}$. During the $(i+1)$th time slot ($1\leq i\leq NL$), the $(i+1)$th column of the $\mathbf{H}_s^\dag$ can be estimated by turning on the $(i+1)$th element of the IRS while keeping all the other elements turned off. The estimates are computed by minimizing the mean square error. Similarly, the primary compositive channel matrices $\{\mathbf{H}_{p,k}\}$ can be estimated by using this protocol, whose details have been omitted for brevity.


\section{Joint Beamforming and Reflecting Coefficient Design for Perfect CSI Case}

\label{sec:perfect channel}
In this section, the perfect CSI case is considered to jointly design the beamforming and reflecting coefficients. Such design will serve as a stepping stone toward developing a more realistic robust design for the case with imperfect CSI. We first formulate the optimization problem and propose an iterative algorithm by applying the BCD method, in which each iteration consists of two subproblems. The first subproblem is equivalently reformulated as an SOCP. The second subproblem is transformed to an SDP by relaxing the rank-1 constraint. Finally, we analyze the convergence and computational complexity of the proposed algorithm.

\subsection{Problem Formulation and Relaxation}
We focus on maximizing the achievable rate of SU-RX, i.e., $\log(1+{\sf SNR}_s)$, by jointly optimizing the reflecting coefficient vector of IRSs and the transmit beamforming vector of SU-TX, namely ${\bm \theta}$ and $\mathbf{w}$, respectively. Additionally, we limit the IT, i.e., ${\sf IT}_{p,k}$, on PU-RXs and the transmit power of SU-TX. The optimization problem is formulated as follows:
\beqi\label{1ob}
(\mathsf{P1})\quad &~\underset{{\bm \theta},\mathbf{w}}{\max}~&\log\left(1+\frac{|{\bm \theta}^\dag\mathbf{H}_s\mathbf{w}|^2}{\sigma_{s}^2}\right)\inum \label{ob1}\\
&{\rm s.t.} &\left|{\bm \theta}^\dag\mathbf{H}_{p,k}\mathbf{w}\right|^2\leq\Gamma_k, ~\forall k\in\mathcal{K},\inum\label{1it}\\
&&|[{\bm \theta}]_l|\leq1,\forall l\in\mathcal{L},\inum\label{Amc}\\
&&\|\mathbf{w}\|^2\leq P,\inum\label{condW}
\eeqi
where $\mathcal{L}=\{1,\ldots,NL+1\}$. In this problem, \eqref{ob1} maximizes the achievable rate of SU-RX; \eqref{1it} is the IT constraint with the interference threshold $\Gamma_k$ for PU-RX $k$; \eqref{Amc} is the passive IRS-gain constraint; and \eqref{condW} is the transmit power constraint with the maximum transmit power $P$ in the SU-TX. In \eqref{condW}, note that $\|{\bf w}\|^2$ is the transmit power of PU-TX, i.e., ${\rm E}\left[\|\mathbf{x}_s\|^2\right] =\|\mathbf{w}\|^2$. Because $\log(x)$ is a monotonically increasing function of $x$, the objective function in \eqref{ob1} is equivalent to the received signal power of PU-RX, i.e., $|{\bm \theta}^\dag\mathbf{H}_s\mathbf{w}|^2$, in the optimization. It is worth noting that the optimization problem is non-convex because the objective function is non-concave over the coupled $\mathbf{w}$ and ${\bm \theta}$, which cannot be easily solved by using a typical convex optimization method. Therefore, to solve (${\sf P1}$), a BCD method is exploited in the next subsection.


\subsection{Alternating Optimization Algorithm Based on BCD}\label{sec:beamforming vector1}
The original problem (${\sf P1}$) can be decoupled into two subproblems, namely (${\sf P1.1}$) and (${\sf P1.2}$), which are described as follows.
\subsubsection{Transmit Beamforming Optimization}
For a given $\bm \theta$, the optimal beamforming vector can be obtained by solving the problem below:
\beqi\label{2ob}
(\mathsf{P1.1})\quad &~\underset{\mathbf{w}}{\max}~&\left|{\bm \theta}^\dag\mathbf{H}_s\mathbf{w}\right|^2 \inum\label{ob2}\\
&{\rm s.t.} ~& \eqref{1it},\eqref{condW}\inum.
\eeqi
It should be noted that problem (${\sf P1.1}$) is nonconvex because the objective function is a nonconcave function of $\mathbf{w}$. This problem however has been widely studied in traditional CR systems. Following \cite{zhang2008exploiting}, if $\mathbf{w}$ is a feasible solution of problem (${\sf P1.1}$), then $e^{j\theta}\mathbf{w}$ for arbitrary $\theta$ is also a feasible solution which maintains the same objective value. Thus, problem (${\sf P1.1}$) can be rewritten as an SOCP as follows:
\beqi\label{P3}
(\mathsf{P1.1'})\quad &~\underset{\mathbf{w}}{\max}~&\mathrm{Re}\left({\bm \theta}^\dag\mathbf{H}_s\mathbf{w}\right) \inum\label{ob3}\\
&{\rm s.t.}&\mathrm{Im}\left({\bm \theta}^\dag\mathbf{H}_s\mathbf{w}\right)=0,\inum\\
~&& \eqref{1it},\eqref{condW}\nonumber.
\eeqi
Subproblem (${\sf P1.1'}$) is convex and can be solved by using existing optimization softwares, such as CVX solvers.
\subsubsection{Reflecting Coefficients Optimization}\label{phase shift}
For a given $\mathbf{w}$, by introducing a new variable ${\bm \Theta}\triangleq{\bm \theta}{\bm \theta}^\dag$, which belongs to rank-$1$ symmetric PSD matrix, problem (${\sf P1}$) can be equivalently rewritten as follows:
\beqi
(\mathsf{P1.2})\quad&~\underset{\bm \Theta}{\max}~&\mathrm{tr}\left({\bm \Theta}\mathbf{H}_s\mathbf{w}\mathbf{w}^\dag\mathbf{H}_s^\dag\right)\inum\label{1it40}\\
& {\rm s.t.} &\mathrm{tr}\left({\bm\Theta}\mathbf{H}_{p,k}\mathbf{w}\mathbf{w}^\dag\mathbf{H}_{p,k}^\dag\right)\leq\Gamma_k,~\forall k\in\mathcal{K},~~~~~~\inum\label{1it41}\\
& &\left[{\bm \Theta}\right]_{l,l}\leq 1,~\forall l\in\mathcal{L},\inum\label{1it42}\\
& &{\bm \Theta}\succeq 0,\inum\label{1it43}\\
& &{\rm rank}({\bm \Theta})=1.\inum\label{R1const}
\eeqi
Note that the objective function in \eqref{1it40} and constraints \eqref{1it41} and \eqref{1it42} are linear, and the symmetric PSD matrix in \eqref{1it43} belongs to a convex set. The rank-$1$ constraint in \eqref{R1const}, however, is non-convex; therefore, we apply the SDR technique to relax this constraint. Consequently, problem ($\sf{P1.2}$) can be compactly rewritten as follows:
\beqi\label{P5}
(\mathsf{P1.2'})\quad&~\underset{\bm \Theta}{\max}~&\mathrm{tr}\left({\bm \Theta}\mathbf{H}_s^\dag\mathbf{w}\mathbf{w}^\dag\mathbf{H}_s\right)\label{obj1}\\
&{\rm s.t.} &\eqref{1it41},\eqref{1it42},\eqref{1it43}\nonumber.
\eeqi
The variable $\bm \Theta$ can be obtained by solving subproblem (${\sf P1.2'}$), which is a standard convex SDP problem. Hence, this problem can be efficiently solved by using, e.g., the interior-point method \cite{grant2014cvx,MOSEK} via CVX and MOSEK solvers.

\subsubsection{Gaussian Randomization for Rank-1 Condition}\label{GR}
Due to the relaxation of rank-$1$ constraints in (${\sf P1.2'}$), the optimal ${\bm \Theta}$ could be infeasible for the original problem (${\sf P1.2}$). To guarantee the feasibility of the converged solutions, i.e., to construct a rank-$1$ solution, we employ a Gaussian randomization scheme \cite{992139}. Using the converged solution of ${\bm \Theta}$ and random vector ${\bf z}\sim\mathcal{CN}({\bf 0}_{NL},{\bf I}_{NL})$, we generate rank-$1$ candidate solution as follows:
\beqi\label{rs}
\widetilde{{\bm \theta}}&\;=\;&\mathbf{U}\sqrt{\mathbf\Sigma}\bmat {\bf z}/\|{\bf z}\|\\1\emat\in\mathbb{C}^{(NL+1)\times1},
\eeqi
where $\mathbf{U}$ is the left singular matrix of ${\bm \Theta}$ and $\mathbf\Sigma$ is its corresponding singular matrix, whose diagonal elements contain the singular values
The randomized solution, $\widetilde{\bm \theta}$, is tested with \eqref{1it11}--\eqref{1it44} to evaluate the feasibility. After generating and testing $G$ randomized solutions, repeatedly, the optimal randomized solution which is feasible and provides the largest objective value in \eqref{1it00}, equivalently $|\widetilde{\bm \theta}^\dag\mathbf{H}_s\widetilde{\mathbf{w}}|$, is determined and denoted by $\widetilde{\bm \theta}^*$. Finally, the reflecting coefficient design for the $n$th IRS is obtained by $\widetilde{\bm \theta}^{(n),*}=\widetilde{\bm \theta}^{*}[(n-1)L+1:nL-1]=\left[\widetilde{\alpha}^{(n)}_1 e^{j\widetilde{\theta}_1^{(n)}},\ldots,\widetilde{\alpha}^{(n)}_L e^{j\widetilde{\theta}^{(n)}_L}\right]$.

In practice, since the reflecting elements shift the phase $\widetilde{\theta}^{(n)}_l$ within the discrete values due to the hardware limitation, a uniform quantization of the phase $\psi\in\mathcal{Q}=\left\{0,2\pi\frac{1}{Q},2\pi\frac{2}{Q},\ldots,2\pi\frac{Q-1}{Q}\right\}$ with $Q$ levels, i.e., a $\log_2(Q)$-bit uniform quantizer, is considered as follows:
\beq \label{des}
 \overline{\theta}^{(n)}_l = \arg \underset{\psi\in\mathcal{Q}}\min\left|{\widetilde{ \theta}^{(n)}_l}-\psi\right|,~\forall l\in\mathcal{L}.
\eeq
Using \eqref{des}, the reflecting coefficient vector of IRS $n$ is modeled as follows:
\beq\label{thetaQ}
\overline{\bm \theta}^{(n),*}=\left[\widetilde{\alpha}^{(n)}_1 e^{j\overline{\theta}_1^{(n)}},\ldots,\widetilde{\alpha}^{(n)}_L e^{j\overline{\theta}^{(n)}_L}\right].
\eeq


\subsubsection{Overall Algorithm}
The variables  $\mathbf w$ and ${\bm \theta}$ are optimized by alternately solving (${\sf P1.1'}$) and (${\sf P1.2'}$) while keeping the other block of variables fixed, i.e., the BCD method. Here, the obtained solution in each iteration is used as the input for the next iteration. After the iterative algorithm converges, we get the converged solution of (${\sf P1}$), denoted by $\widetilde{\mathbf w}^{*}$ and $\widetilde{\bm \theta}^{*}$. Later, the discrete phase shift values $\overline{\bm \theta}^{(n),*}$ for $n\in\mathcal{N}$ are obtained through quantization process. Algorithm \ref{Alg1} is the summary of the BCD and Gaussian randomization algorithm to achieve $\widetilde{\bf w}^*$ and $\overline{\bm \theta}^{(n),*}$ for a given tolerance factor and a number of random realizations, which are denoted by $\epsilon$ and $G$, respectively.

\begin{algorithm}
\SetKw{to}{to}
\SetKw{fulfills}{fulfills}
\SetKw{IE}{i.e.}
\SetKw{feasible}{feasible}
\caption{BCD Algorithm with Perfect CSI}\label{Alg1}
{\bf Input}: $\mathbf{H}_s$ and $\{\mathbf{H}_{p,k}\}$\\
{\bf Output}: $\widetilde{\mathbf{w}}^*$ and ${\overline{\bm \theta}^{(n),*}}$ for $n\in\mathcal{N}$\\
{Initialize $\bm \theta$ by an all-one matrix, $\eta_a=0$, $\eta_{0}=\epsilon$, $\delta=\epsilon+1$, and $i=0$.}\\
\underline{BCD-based optimization}\\
\While{$\delta > \epsilon$}{
Obtain $\widetilde{\mathbf{w}}_{i+1}$ by solving SOCP (${\sf P1.1'}$) in \eqref{P3} for given $\widetilde{\bm \theta}^*_{i}$.\label{AgLine2}\\
Obtain ${\bm \Theta}_{i+1}$ by solving SDR-SDP (${\sf P1.2'}$) in \eqref{P5} for given $\widetilde{\mathbf{w}}_{i+1}$.\label{AgLine1}\\
\underline{Gaussian randomization}\\
${\bm \Theta}_{i+1}\overset{svd}{=}\mathbf{U}_{i+1}\mathbf\Sigma_{i+1}\mathbf{V}^{H}_{i+1}$\\
 \For{$g=1$ \to $G$}{
    Generate a random vector ${\bf z}\in\mathbb{C}^{(NL+1)\times1}$.\\
    Obtain $\widetilde{{\bm \theta}}_{i+1}$ from \eqref{rs} with ${\bf z}$ and ${\bm \Theta}_{i+1}$.\\
    \If{$\widetilde{{\bm \theta}}_{i+1}$ \fulfills \eqref{1it41}--\eqref{1it43}, \IE, \feasible,}{
       Compute $\eta_b=\left|\widetilde{\bm \theta}_{i+1}^\dag\mathbf{H}_s\widetilde{\mathbf{w}}_{i}\right|^2$.\\
       \If{$\eta_b\geq \eta_a$}{
       Update $\widetilde{\bm \theta}^*_{i+1}=\widetilde{{\bm \theta}}_{i+1}$ and $\eta_a=\eta_b$.
       }
    }
  }
Compute $\eta_{i+1}=\left|\widetilde{\bm \theta}^{*,\dag}_{i+1}\mathbf{H}_s\widetilde{\mathbf{w}}_{i}\right|^2$ and $\delta = \left|\eta_{i+1}-\eta_{i} \right|/ \eta_{i}$.\\
Update $i\leftarrow i+1$
}
Set $\widetilde{\mathbf{w}}^{*}=\widetilde{\mathbf{w}}_{i}$, $\widetilde{\bm \theta}^{*}=\widetilde{\bm \theta}^*_{i}$, and $\widetilde{\bm \theta}^{(n),*}=\widetilde{\bm \theta}^{*}[(n-1)L+1: nL-1]$ for $n\in\mathcal{N}$.\\
Obtain ${\overline{\bm \theta}^{(n),*}}$ from a uniform quantization of phase as shown in \eqref{des} and \eqref{thetaQ}.
\end{algorithm}

The convergence of the whole algorithm is discussed in the following property.

\property \label{Pro0}Algorithm 1 is guaranteed to converge if the variables obtained by solving \eqref{P3} and \eqref{P5} in the $i$th iteration satisfy
\beqi
 \mathrm{SNR_s}\left({\bm \Theta}_{i+1},\mathbf{w}_{i}\right)
 &\;\geq\;&\mathrm{SNR_s}\left({\bm \Theta}_{i},\mathbf{w}_{i}\right)\nonumber\\
 &\geq&\mathrm{SNR_s}({\bm \Theta}_{i},\mathbf{w}_{i-1}),
 \eeqi
 which means that
\beq \label{conver}
 \eta_{i+1}\geq\eta_i.
 \eeq
\begin{IEEEproof}
 It can be verified that the objective function is monotonically nondecreasing after each iteration when \eqref{conver} is satisfied. Therefore, Algorithm 1 is guaranteed to converge.
\end{IEEEproof}

\property \label{ProC1} The time complexity of Algorithm 1 is $\mathcal{O}(T_1(M^2K^{1.5}+M^3K^{0.5}+KN^2L^2M+N^{4.5}L^{4.5}\log(1/\varepsilon)+N^3L^3+GKNL ))$, where $T_1$ is the number of iterations required to converge for a given tolerance factor $\epsilon$ and solution accuracy $\varepsilon$ of the interior-point method.
\begin{IEEEproof}
In each iteration, the complexity for computing ${\bm \theta}^\dag\mathbf{H}_{s}$ and ${\bm \theta}^\dag\mathbf{H}_{p,k}$ for $k\in\mathcal{K}$ is $\mathcal{O}((K+1)(NL+1)M)\simeq \mathcal{O}(KNLM)$. The SOCP in step $6$ can be solved efficiently using a primal-dual interior-point method with the worst-case complexity of $\mathcal{O}(M^2(2K+M+1)(K+1)^{0.5})\simeq \mathcal{O}(M^2K^{1.5}+M^3K^{0.5})$.
The complexity for computing  $\mathbf{H}_s^\dag\mathbf{w}\mathbf{w}^\dag\mathbf{H}_s^\dag$ and
$\mathbf{H}_{p,k}\mathbf{w}\mathbf{w}^\dag\mathbf{H}_{p,k}^\dag$ for $k\in\mathcal{K}$ is $\mathcal{O}((K+1)(NL+1)^2M)\simeq \mathcal{O}(KN^2L^2M)$. The SDP in step $7$ is solved by using prominent interior-point algorithm with the worst-case complexity of $\mathcal{O}((NL+1)^{4.5}\log(1/\varepsilon))\simeq \mathcal{O}(N^{4.5}L^{4.5}\log(1/\varepsilon))$.
During the Gaussian randomization, the complexity for calculating singular value decomposition (SVD) of $\bm \Theta$ is given by $\mathcal{O}((NL+1)^3)\simeq \mathcal{O}(N^3L^3)$. In steps $15$ and $16$, the complexity of checking the feasibility and calculating $\eta_b$ is $\mathcal{O}((K+1)(NL+1))\simeq\mathcal{O}(KNL)$.
Thus, the overall time complexity of Algorithm $1$ is obtained as shown in Property \ref{ProC1}.
\end{IEEEproof}

\section{Robust Joint Beamforming and Phase Shift Design for Imperfect CSI Case}
\label{sec:imperfect channel}

The imperfect CSI case is studied in this section.
In this study, the uncertain CSI is modeled by an ellipsoid uncertainty region \cite{zheng2009robust,wang2009worst}, and robust joint beamforming and phase shifts against are designed the imperfect CSI. With the channel uncertainty, the original problem (${\sf P1}$) is equivalently modeled as a worst-case minmax problem. Such problem however contains an infinite number of constraints, thus it cannot be directly solved. By using a heuristic transform and BCD, we propose an alternative algorithm in which the intractable problem is reformulated into the tractable SDP problems. The convergence and computational complexity of the proposed algorithm are also analyzed.

\subsection{Problem Formulation with CSI Uncertainty}\label{sec:beamforming vector}
Similar to the variable $\bm \Theta$ in \ref{phase shift}, we introduce another new variable $\mathbf{W}\triangleq\mathbf{w}\mathbf{w}^\dag$, which is a rank-$1$ symmetric PSD matrix. Using these two variables, $\bm \Theta$ and $\mathbf{W}$, and applying the SDR technique to relax the rank-$1$ constraints of ${\bf W}$ and ${\bm \Theta}$, problem ($\mathsf{P1}$) can be rewritten as follows:
\beqi
(\mathsf{P1'})\quad&~\underset{{\bm \Theta},\mathbf{W}}{\max}~&\mathrm{tr}\left({\bm \Theta}\mathbf{H}_s\mathbf{W}\mathbf{H}_s^\dag\right)\inum\label{1it00}\\
& {\rm s.t.} &\mathrm{tr}\left({\bm\Theta}\mathbf{H}_{p,k}\mathbf{W}\mathbf{H}_{p,k}^\dag\right)\leq\Gamma_k,~\forall k\in\mathcal{K},\inum\label{1it11}\\
& &[{\bm \Theta}]_{l,l}\leq 1,~\forall l\in\mathcal{L},\inum\label{1it22}\\
& &\mathrm{tr}(\mathbf{W})\leq P,\inum\label{1it33}\\
& &{\bm \Theta}\succeq 0,\inum\label{1it44}\\
& &\mathbf{W}\succeq 0.\inum\label{1it55}
\eeqi
\newcounter{mytempeqncnt}

To model the uncertainty of the channel deterministically \cite{zhang2009robust,cumanan2008robust,pei2011secure,zhang2012distributed,wang2014robust}, we assume that the perfect channels, $\mathbf{H}_s$ and $\{\mathbf{H}_{p,k}\}$, lie in the neighborhood of the corresponding estimated channels, which are known at the SU-TX as follows:
\beqi\label{CSIerr}
\mathbf{H}_s&\;=\;&\hat{\mathbf{H}}_s + \Delta_{s}\in\mathbb{C}^{(NL+1)\times M},\inum\label{channel1}\\
\mathbf{H}_{p,k}&=&\hat{\mathbf{H}}_{p,k} +\Delta_{p,k}\in\mathbb{C}^{(NL+1)\times M},\inum\label{channel2}
\eeqi
for $k\in\mathcal{K}$, where $\hat{\mathbf{H}}_s$ and $\hat{\mathbf{H}}_{p,k}$ denote the estimated channels known at the SU-TX, and $\Delta_{s}$ and $\{\Delta_{p,k}\}$ denote the CSI errors. The uncertain CSI is bounded in a set defined by the following ellipsoid uncertainty region \cite{zheng2009robust,wang2009worst}:
\beqi\label{CSIDel}
\mathcal{H}_s &\;=\;& \left\{ \Delta_{s}:~{\rm tr}\left(\Delta_{s}{\bf C}_s\Delta_s^\dag\right) \leq 1 \right\},\inum\label{CSI1}\\
\mathcal{H}_{p,k}&=&\left\{\Delta_{p,k}:~{\rm tr}\left(\Delta_{p,k}{\bf C}_{p,k}\Delta_{p,k}^\dag\right) \leq 1\right\},\inum\label{CSI2}
\eeqi
for $k\in\mathcal{K}$, where ${\bf C}_s\succ0$ and ${\bf C}_{p,k}\succ0$ are the scaled inverse covariance matrices of ${\bf H}_s$ and ${\bf H}_{p,k}$, respectively, which determine the boundary of the uncertainty region. Note that, in the deterministic model, $\left\{{\bf C}_s\in\mathbb{R}^{M\times M},{\bf C}_{p,k}\in\mathbb{R}^{M\times M}\right\}$ are known at the PU-TX.


Considering the channel errors in \eqref{CSIerr}, the robust beamforming and phase shifts within the deterministic uncertainty region can be designed by solving a minmax problem, which is modified from problem (${\sf P1'}$), as follows:
\beqi
(\mathsf{P2})\quad&\underset{{\bm \Theta},\mathbf{W}}{\max}~\underset{\Delta_{s}}{\min}\;&
\mathrm{tr}\left({\bm \Theta}\mathbf{H}_s\mathbf{W}\mathbf{H}_s^\dag\right)\inum\label{objective}\\
&{\rm s.t.}& \Delta_{s}\in\mathcal{H}_s,\inum\label{erCon1}\\
&&\Delta_{p,k}\in\mathcal{H}_{p,k},~\forall k\in\mathcal{K},\inum\label{erCon2}\\
&& \eqref{1it11},\eqref{1it22},\eqref{1it33},\eqref{1it44},\eqref{1it55}.\nonumber
\eeqi
where the objective function in \eqref{objective} is the worst-case (i.e., minimum) received SNR at the SU-RX.

We observe that the objective function in \eqref{objective} is biconvex in ${\bm \Theta}$ and $\mathbf{W}$ for fixed $\Delta_{s}$, and concave in $\Delta_{s}$ for fixed  ${\bm \Theta}$ and $\mathbf{W}$. By introducing an auxiliary variable $t$, the minmax problem (${\sf P2}$) can be reformulated as the following maximization problem:
\beqi
(\mathsf{P2'})\quad&\;\underset{{\bm \Theta},\mathbf{W},t}{\max}\; &t\inum\\
&{\rm s.t.}&  \mathrm{tr}\left({\bm \Theta}\mathbf{H}_s\mathbf{W}\mathbf{H}_s^\dag\right) \geq t \inum\label{aux}\\
&&\eqref{1it11},\eqref{1it22},\eqref{1it33},\eqref{1it44},\eqref{1it55},\eqref{erCon1},\eqref{erCon2}.\nonumber
\eeqi

Problem (${\sf P2'}$) is a mixed-integer non-convex problem, whose optimal solution cannot be obtained easily. Therefore, following a similar procedure mentioned in Subsection \ref{sec:beamforming vector1}, we adopt the BCD technique to optimize $\mathbf{\Theta}$ and ${\bm W}$ alternately, and eventually obtain an efficient sub-optimal solution of (${\sf P2'}$).

\subsection{Alternating Optimization Based on BCD}\label{BCD2}

Based on BCD, (${\sf P2'}$) can be decoupled into two subproblems (${\sf P2'.1}$) and (${\sf P2'.2}$) for given $\mathbf{W}$ and ${\bm \Theta}$, respectively, as follows:
\beq
\begin{split}
({\sf P2'.1})\quad~\underset{{\bm \Theta},t}{\max}\;& t\inum \\
{\sf s.t.}~&\eqref{1it11},\eqref{1it22},\eqref{1it44},
 \eqref{erCon1},\eqref{erCon2},\eqref{aux},\label{P7.1}
\end{split}
\eeq
and
\beq
\begin{split}
({\sf P2'.2})\quad~\underset{\mathbf{W},t}{\max}\;& t\\
{\sf s.t.}~& \eqref{1it11},\eqref{1it33},\eqref{1it55},\eqref{erCon1},\eqref{erCon2},\eqref{aux}.\label{P7.2}
\end{split}
\eeq

The quadratic optimization problems, (${\sf P2'.1}$) and (${\sf P2'.2}$) are called linear semi-infinite programming owing to the infinite number of constraint sets with finite number of variables, which is caused by the channel uncertainty. To efficiently solve (${\sf P2'.1}$) and (${\sf P2'.2}$), we reformulate each subproblem as an SDP problem. To accomplish this, we briefly introduce a powerful tool, called $\mathcal{S}$-Procedure \cite{beck2006strong}, and use it to reformulate the infinite number of constraints.
\begin{figure*}[!b]
\vspace*{4pt}\hrulefill%
\normalsize \setcounter{mytempeqncnt}{\value{equation}} %
\setcounter{equation}{29}%
\beqi
\bmat
\rho_s\left(\mathbf{C}_{s}^\dag\otimes\mathbf I_M\right)+\mathbf{W}^\dag\otimes{\bm \Theta} & \mathrm{vec}\left(\mathbf{W}\hat{\mathbf{H}}_s^\dag{\bm \Theta}\right)   \\
\mathrm{vec}^\dag\left(\mathbf{W}\hat{\mathbf{H}}_s^\dag{\bm \Theta}\right) &-\rho_s+\mathrm{tr}\left({\bm \Theta}\hat{\mathbf{H}}_s\mathbf{W}\hat{\mathbf{H}}_s^\dag\right)-t\\
\emat
&\;\succeq\;& 0,\inum\label{matrix_s1}\\
\bmat
\rho_{p,k}\left(\mathbf{C}_{p,k}^\dag\otimes\mathbf I_M\right)& -\mathbf{W}^\dag\otimes{\bm \Theta}-\mathrm{vec}\left(\mathbf{W}\hat{\mathbf{H}}_{p,k}^\dag{\bm \Theta}\right)   \\
-\mathrm{vec}^\dag\left(\mathbf{W}\hat{\mathbf{H}}_{p,k}^\dag{\bm \Theta}\right) &-\rho_{p,k}-\mathrm{tr}\left({\bm \Theta}\hat{\mathbf{H}}_{p,k}\mathbf{W}\hat{\mathbf{H}}_{p,k}^\dag\right)+\Gamma_k\\
\emat
&\succeq &0,~\forall k\in\mathcal{K}.\inum\label{matrix_p1}
\eeqi
\setcounter{equation}{\value{mytempeqncnt}}
\end{figure*}

{\it $\mathcal{S}$-Procedure:
Let $f_k(\mathbf x)$, $k=\{1, 2\}$, be
\begin{equation}
f_k({\bf x})=\mathbf x^\dag\mathbf A_k\mathbf x+2\mathrm{Re}\left\{{\bf b}_k^\dag {\bf x}\right\}+c_k,
\end{equation}
where $\mathbf A_k$ is an $N\times N$ Hermitian matrix, $\bf x$ and $\bf b_k$ are $N\times1$ complex vectors, and $c_k$ is a real scalar.
Assume that there exists $\hat{\mathbf x}$ such that $f_1\left(\hat{\mathbf x}\right)<0$. Then, the implication $f_1({\bf x})\leq0\Rightarrow f_2({\bf x})\leq0$ holds true if and only if there exists $\rho\geq0$ such that}
\begin{equation}
  \rho
\begin{bmatrix}
\mathbf A_1  &   \mathbf b_1    \\
\mathbf b_1^\dag&  c_1\\
\end{bmatrix}-
\begin{bmatrix}
\mathbf A_2  &   \mathbf b_2    \\
\mathbf b_2^\dag&  c_2\\
\end{bmatrix}\succeq 0.
\end{equation}

The following properties are obtained by using the $\mathcal{S}$-Procedure.

\property \label{Pro1}Constraints \eqref{erCon1} and \eqref{aux}, i.e., the SNR constraints with CSI uncertainty, are equivalent to \eqref{matrix_s1} at the bottom of this page. Here, $\rho_s\geq0$.

\addtocounter{equation}{1}
\begin{IEEEproof}
According to the assumption of CSI errors in \eqref{CSIDel}, the uncertainty constraint \eqref{erCon1} can be readily rewritten as follows:

\beq
\mathrm{vec}^\dag\left(\Delta_s\right)\left(\mathbf{C}_{s}^\dag\otimes\mathbf I_M\right)\mathrm{vec}(\Delta_s) \leq 1.\label{S2}
\eeq
In contrast, the robust constraint in \eqref{aux} can further be derived by using \eqref{channel1} as follows:
\beqi
&\mathrm{tr}&\left[{\bm \Theta}\left(\hat{\mathbf{H}}_s+\Delta_s\right)\mathbf{W}\left(\hat{\mathbf{H}}_s+\Delta_s\right)^\dag\right] \geq t\nonumber\\
&&\Leftrightarrow\mathrm{tr}\Big[{\bm \Theta}\Delta_s\mathbf{W}\Delta_s^\dag+{\bm \Theta}\hat{\mathbf{H}}_s\mathbf{W}\Delta_s^\dag+{\bm \Theta}\Delta_s\mathbf{W}\hat{\mathbf{H}}_s^\dag\nonumber\\
&&\quad\quad\quad+{\bm \Theta}\hat{\mathbf{H}}_s\mathbf{W}\hat{\mathbf{H}}_s^\dag\Big]\geq t\nonumber\\
&&\Leftrightarrow \mathrm{vec}^\dag(\Delta_s)\left(\mathbf{W}^\dag\otimes{\bm \Theta}\right)\mathrm{vec}(\Delta_s)\nonumber\nonumber\\
&&\quad\quad\quad+2\mathrm{Re}\left\{\mathrm{vec}^\dag\left(\mathbf{W}\hat{\mathbf{H}}_s^\dag{\bm \Theta}\right)\mathrm{vec}\left(\Delta_s^\dag\right)\right\}\nonumber\\
&&\quad\quad\quad\quad+\mathrm{tr}\left[{\bm \Theta}\hat{\mathbf{H}}_s\mathbf{W}\hat{\mathbf{H}}_s^\dag\right]\geq t.\label{S1}
\eeqi
By using $\mathcal{S}$-procedure, we can show that \eqref{S2} and \eqref{S1} are valid if and only if there exists $\rho_s\geq 0$ such that \eqref{matrix_s1} is valid. This completes the proof.
\end{IEEEproof}

\property \label{Pro2}Constraints \eqref{1it11} and \eqref{erCon2}, i.e., the IT constraints with CSI uncertainty, are equivalent to \eqref{matrix_p1} at the bottom of this page.

\begin{IEEEproof}
According to the assumption of CSI errors in \eqref{CSIDel}, the uncertainty constraints, namely \eqref{erCon1} and \eqref{erCon2}, can be readily rewritten as follows:
\beq
\mathrm{vec}^\dag(\Delta_{p,k})\left(\mathbf{C}_{p,k}^\dag\otimes \mathbf I_M\right)\mathrm{vec}(\Delta_{p,k})\leq 1.\label{S3}
\eeq
Similarly, the IT constraint \eqref{1it11} can be further written as follows:
\beqi\label{S-1}
&&\mathrm{vec}^\dag\left(\Delta_{p,k}\right)\left(\mathbf{W}^\dag\otimes{\bm \Theta}\right)\mathrm{vec}(\Delta_{p,k})\nonumber\\
&&\quad+2\mathrm{Re}\left\{\mathrm{vec}^\dag\left(\mathbf{W}\hat{\mathbf{H}}_{p,k}^\dag{\bm \Theta}\right)\mathrm{vec}\left(\Delta_{p,k}^\dag\right)\right\}\nonumber\\
&&\qquad+\mathrm{tr}\left[{\bm \Theta}\hat{\mathbf{H}}_{p,k}\mathbf{W}\hat{\mathbf{H}}_{p,k}^\dag\right]\geq\Gamma_k,
\eeqi
for $k\in\mathcal{K}$. Using $\mathcal{S}$-Procedure again, we can show that \eqref{S-1} and \eqref{S3} are valid if and only if there exists $\rho_{p,k}\geq0$ such that \eqref{matrix_p1} is valid. This completes the proof.
\end{IEEEproof}

Using Lemmas 1 and 2, we now recast the infeasible semi-infinite programming subproblems, namely (${\sf P2'.1}$) and (${\sf P2'.2}$), to the tractable convex SDP problems as follows:
\beq
\begin{split}
({\sf P2'.1'})\quad~&\underset{\rho_s\geq0,\{\rho_{p,k}\geq0\},{\bm \Theta},t}{\max} ~ t\inum \\
&{\sf s.t.}~\eqref{1it22},\eqref{1it44},\eqref{matrix_s1}, \eqref{matrix_p1},
\label{P7.1p}
\end{split}
\eeq
and
\beq
\begin{split}
({\sf P2'.2'})\quad~&\underset{\rho_s\geq0,\{\rho_{p,k}\geq0\},\mathbf{W},t}{\max} ~ t\\
&{\sf s.t.}~ \eqref{1it33},\eqref{1it55},\eqref{matrix_s1},\eqref{matrix_p1}.\label{P7.2p}
\end{split}
\eeq

It is worth noting that (${\sf P2'.1'}$) and (${\sf P2'.2'}$) are now the standard convex SDP problems and hence, they can be readily solved by using a standard interior-point method. Therefore, using the BCD method, the variables $\mathbf \Theta$ and ${\bm W}$ are optimized alternately by solving (${\sf P2'.1'}$) and (${\sf P2'.2'}$), respectively. After obtaining the converged solutions $\mathbf \Theta^{*}$ and ${\bm W^{*}}$, similar to Algorithm 1, we can apply the Gaussian randomization scheme to find the generally approximate solution $\widetilde{{\bm \theta}}^*$ and $\widetilde{\mathbf{w}}^*$. The description of the procedures is omitted here for brevity, and the overall algorithm is summarized in Algorithm 2.

\property \label{ProC2} The time complexity of Algorithm 2 is $\mathcal{O}(T_2((N^{4.5}L^{4.5}+M)\log(1/\varepsilon)+KN^2L^2M )+N^3L^3+M^3+GKNLM)$, where $T_2$ is the number of iterations required for the convergence of the BCD procedure with a given tolerance factor $\epsilon$ and accuracy $\varepsilon$ of an interior-point method.
\begin{IEEEproof}
For ease of analysis, without loss of generality, the CSI uncertainty is ignored in the complexity analysis. The complexity for computing $\mathbf{H}_s\mathbf{W}\mathbf{H}_s^\dag$ and $\mathbf{H}_{p,k}\mathbf{W}\mathbf{H}_{p,k}^\dag$ for $k\in\mathcal{K}$ is $\mathcal{O}((K+1)(NL+1)^2M)\simeq\mathcal{O}(KN^2L^2M)$.
The worst-case complexity of the first SDP subproblem is $\mathcal{O}((NL+1)^{4.5}\log(1/\varepsilon))\simeq\mathcal{O}(N^{4.5}L^{4.5}\log(1/\varepsilon))$. The complexity for computing $\mathbf{H}_s^\dag \mathbf{\Theta} \mathbf{H}_s$ and $\mathbf{H}_{p,k}^\dag\mathbf{\Theta}\mathbf{H}_{p,k}$ for $k\in\mathcal{K}$ is $\mathcal{O}((K+1)(NL+1)^2M)\simeq\mathcal{O}(KN^2L^2M)$. The worst-case complexity of the second SDP subproblem is $\mathcal{O}(M^{4.5}\log(1/\varepsilon))$.
During the Gaussian randomization process, the complexity for calculating SVD of $\bm \Theta$ and ${\bf W}$ is $\mathcal{O}( (NL+1)^3+ M^3)\simeq\mathcal{O}(N^3L^3+M^3)$, and the complexity in step $17$ and step $18$ is $\mathcal{O}((K+1)(NL+1)M)\simeq\mathcal{O}(KNLM)$.
Thus, the overall time complexity of Algorithm $2$ is obtained as shown in Property \ref{ProC2}.
\end{IEEEproof}
\begin{algorithm}
\SetKw{to}{to}
\SetKw{fulfill}{fulfill}
\SetKw{IE}{i.e.}
\SetKw{and}{and}
\SetKw{feasible}{feasible}
\caption{BCD Algorithm with Imperfect CSI}\label{Alg2}
{\bf Input}: $\hat{\mathbf{H}}_s$ and $\left\{\hat{\mathbf{H}}_{p,k}\right\}$\\
{\bf Output}: $\widetilde{\mathbf{w}}^*$ and ${\overline{\bm \theta}^{(n),*}}$ for $n\in\mathcal{N}$\\
{Initialize $\mathbf{W}$ by an all-one matrix, $\eta_a=0$, $\eta_0=\epsilon$, $\delta=\epsilon+1$, and $i=0$.}\\
\underline{BCD-based optimization}\\
\While{$\delta>\epsilon$}{
Obtain ${\bm \Theta}_{i+1}$ by solving SDR-SDP (${\sf P2'.1'}$) in \eqref{P7.1p} for given $\mathbf{W}_{i}$.\\
Obtain $\mathbf{W}_{i+1}$\! by solving SDR-SDP (${\sf P2'.2'}$) in \eqref{P7.2p} for given ${\bm \Theta}_{i+1}$.\\
Compute $\eta_{i+1}=\mathrm{tr}\left({\bm \Theta}_{i+1}\mathbf{H}_s\mathbf{W}_{i+1}\mathbf{H}_s^\dag\right)$ and $\delta = \left| \eta_{i+1}-\eta_{i}\right|/\eta_{i}$.\\
Update $i\leftarrow i+1$
}
Set ${\bm \Theta}^{*}={\bm \Theta}_{i+1}$ and $\mathbf{W}^{*}=\mathbf{W}_{i+1}$.\\
\underline{Gaussian randomization}\\
${\bm \Theta}^{*}\overset{svd}{=}\mathbf{U}_{\Theta}\mathbf\Sigma_{\Theta}\mathbf{V}_{\Theta}^{H}$\\
$\mathbf{W}^{*}\overset{svd}{=}\mathbf{U}_W\mathbf\Sigma_W\mathbf{V}_W^{H}$\\
\For{$g=1$ \to $G$}{
    Generate random vectors ${\bf z}_1\in\mathbb{C}^{(NL+1)\times1}$ and ${\bf z}_2\in\mathbb{C}^{M\times1}$.\\
    Obtain $\widetilde{{\bm \theta}}$ from \eqref{rs} with ${\bf z}_1$ and ${\bm \Theta}^{*}$.\\
    Similarly, obtain $\widetilde{\mathbf{w}}$ from ${\bf z}_2$ and $\mathbf{W}^{*}$.\\
    \If{$\widetilde{{\bm \theta}}$ \and $\widetilde{\mathbf{w}}$ \fulfill \eqref{1it11}--\eqref{1it55}, \IE, \feasible,}{
       Compute $\eta_b=\left|\widetilde{\bm \theta}^\dag\mathbf{H}_s\widetilde{\mathbf{w}}\right|$.\\
       \If{$\eta_b\geq \eta_a$}{
       Update $\widetilde{{\bm \theta}}^*=\widetilde{{\bm \theta}}$ and $\widetilde{\mathbf{w}}^*=\widetilde{\mathbf{w}}$.}{
       Update $\eta_a=\eta_b$.
       }
    }
 }
 Set $\widetilde{\bm \theta}^{(n),*}=\widetilde{\bm \theta}^{*}[(n-1)L+1:nL-1]$ for $n\in\mathcal{N}$.\\
Obtain ${\overline{\bm \theta}^{(n),*}}$ from \eqref{des} and \eqref{thetaQ}.
\end{algorithm}

\section{Simulations Results and Discussions}\label{sec:simulations}
In this section, simulation results are provided to evaluate the performance of the proposed IRS-assisted CR system. We first show the computational complexity and convergence of the proposed algorithms, then examine the achievable rate of the proposed IRS-assisted downlink MISO CR system. The cases for both perfect and imperfect CSI are simulated.

\begin{figure}[!t]
\centering
    \includegraphics[width=0.99\columnwidth]{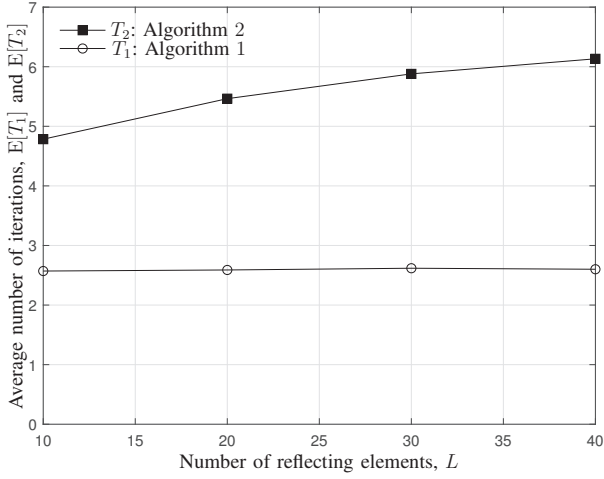}
  \caption{Average number of iterations for BCD-based optimization, namely ${\rm E}[T_1]$ and ${\rm E}[T_2]$ for Algorithms $1$ and $2$, respectively, over $L$.}\label{fig:iterations}
\end{figure}

In Fig. \ref{fig:iterations}, the average numbers of iterations for BCD-based optimization in Algorithms $1$ and $2$, namely ${\rm E}[T_1]$ and ${\rm E}[T_2]$, are shown over number of reflecting elements $L$, when $M=4$, $K=2$, $N=2$, $\varepsilon=10^{-2}$, $\epsilon=10^{-4}$, $P=10\;{\rm dB}$, and $\Gamma=5\;{\rm dB}$. It is observed that Algorithms $1$ and $2$ converge with less than seven iterations. From the results, we can conclude that the proposed algorithms converse generally fast.

By setting $T_1$ and $T_2$ as the average values of ${\rm E}[T_1]$ and ${\rm E}[T_2]$, respectively, for a given $L$, the computational complexity of Algorithms $1$ and $2$ is evaluated numerically in terms of run time. Specifically, from the results in Fig. \ref{fig:iterations}, we set the parameters as follows: $T_1 = 2.5$ for all $L$, while $T_2$ is set to be $4.78$, $5.46$, $5.88$ and $6.29$ when $L$ is $10$, $20$, $30$, and $40$, respectively. Simulations are performed using a server with i7-3770 $3.6\;{\rm GHz}$ CPU, $32\;{\rm GB}$ RAM, and $64$-bit operating system. To verify the complexity analyses in Properties $2$ and $5$, in Fig. 3, we depict the numerical run time with respect to number of reflecting elements $L$. From the results, it is observed that the trends of the complexity analysis and numerical run time match well with each other. In this example simulation, it is also shown that the complexity of Algorithm $2$ is higher than that of Algorithm $1$ due to the fact that solving SOCP (${\sf P1.1'}$) in line $6$ of Algorithm $1$ by an interior-point method requires less complexity than solving SDP (${\sf P2'.1'}$) in line $6$ of Algorithm $2$. Note that although the Gaussian randomization is performed in each BCD iteration in Algorithm $1$, the required computational complexity is relatively lower than the complexity gap between SOCP and SDP.

\begin{figure}[!t]
\centering
    \includegraphics[width=0.99\columnwidth]{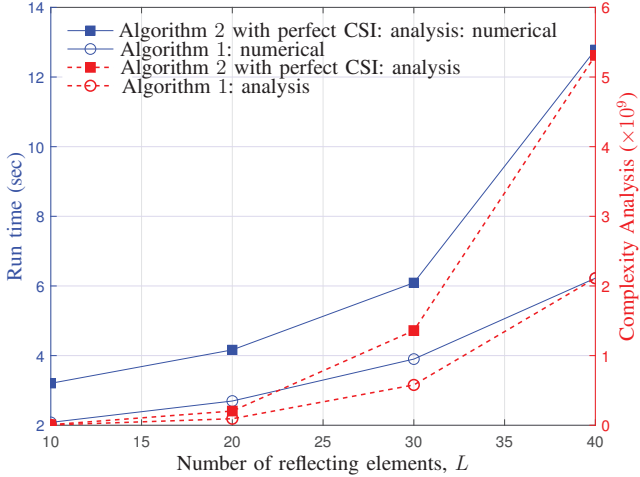}
  \caption{Complexity comparison of Algorithms $1$ and $2$.}\label{fig:complexity}
\end{figure}

In the achievable rate evaluation, an IRS-assisted CR system is set up with a four-antenna SU-TX, single-antenna SU-RX, and two single-antenna PU-RXs, i.e., $M=4$ and $K=2$. Unless otherwise stated, two IRSs are deployed to assist the CR system, i.e., $N=2$. The variances of the AWGN noise at the SU-RX and PU-RXs are fixed to one. The IRSs are deployed more closer to the BS and SU-RX in comparison to the PU-RX; therefore, the Rician factors are set as follows: $\kappa_1=\kappa_3=10$ and $\kappa_2=1$. The relative channel gain is set to be $\varphi=-10\;\rm dB$ \cite{guo2019weighted}. In addition, we assume that all PU-RXs have the same IT limits, i.e. $\Gamma_k=\Gamma,~\forall k$. The maximum transmit power and the IT limits are defined in the ${\rm dB}$ scale with respect to the noise power. 

\begin{figure}[!t]
\centering
    \includegraphics[width=0.99\columnwidth]{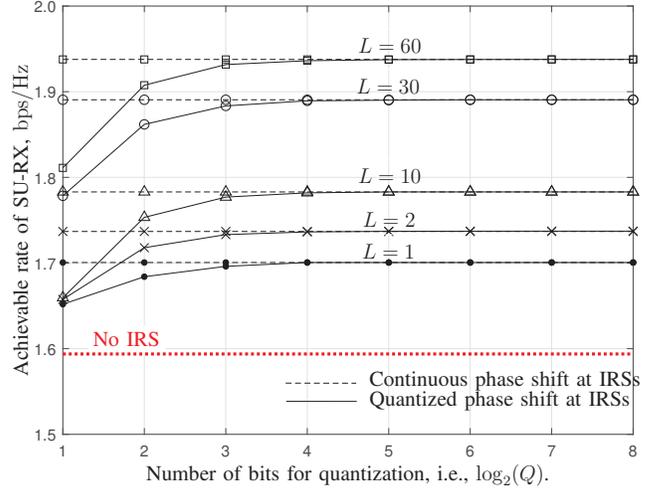}
  \caption{Achievable rate performance over the quantization levels, $Q$, of phase shift at IRSs when $N=2$, $\Gamma=5\;{\rm dB}$, and $P=6\;{\rm dB}$.}\label{fig:quantization}
\end{figure}

In Fig. \ref{fig:quantization}, the achievable rate of the proposed IRS-assisted CR systems is evaluated across the number of quantization bits, i.e., $\log_2(Q)$, for various $L\in\{10,30,60\}$, when $N=2$, $P=6\;{\rm dB}$, and $\Gamma=5\;{\rm dB}$. From the results, we observe that five bits are sufficient for the uniform quantization of phase shift at the IRSs. Therefore, we set $Q=2^5$ for the uniform quantization of the discrete phase shifts at IRSs in simulation. It is, however, worth emphasizing that the proposed IRSs can improve the achievable rate even with an IRS having a single element and one-bit phase shift quantizer.

\subsection{Perfect CSI Case}
In the perfect CSI case, the proposed IRS-assisted CR system is compared with two benchmark schemes, as given below.
\begin{itemize}
\item CR with Random IRS: The reflecting coefficients of each element of each IRS are randomly selected, i.e., $\theta_l^{(n)}\sim\mathcal{U}(0,2\pi),~\forall l\in\mathcal{L}$ and $\alpha^{(n)}_l=1,~\forall n$ and $\forall l$. Under this setup, only problem (${\sf P2.1'}$) requires a solution to obtain the SU-TX beamforming vector ${\bf w}$, which is a typical SOCP problem and can be optimally solved.
\item CR without IRS: This system is a traditional downlink MISO CR system without IRS. The optimal beamforming vector ${\bf w}$ of SU-RX in a traditional CR system can be obtained from (${\sf P2.1'}$) by setting the reflect-link channels to be null, namely $\mathbf{F}=\mathbf{0}$, $\mathbf{h}_{r,s}=\mathbf{0}$, and $\mathbf{h}_{r,p,k}=\mathbf{0}$.
\end{itemize}

\begin{figure}[!t]
\centering
    \includegraphics[width=0.99\columnwidth]{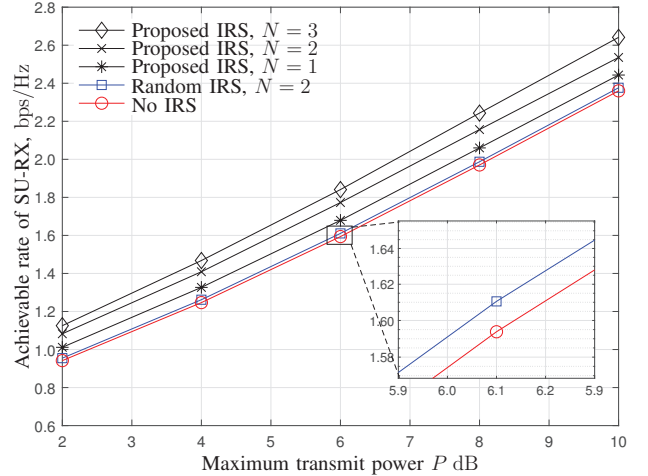}
  \caption{Achievable rate of SU-RX versus the maximum transmit power of SU-TX, $P$, when $\Gamma=5\;{\rm dB}$, $L=10$ and $N=1$, $2$, and $3$.}\label{perfect}
\end{figure}

Fig. \ref{perfect} shows the achievable rate as a function of the maximum transmit power $P$ when the IT tolerance level is $\Gamma=5\;{\rm dB}$. Different numbers of IRSs are considered, i.e., $N\in\{1,2,3\}$, and the number of each IRS element is set to ten, i.e., $L=10$. As expected, the achievable rates of all schemes increase with $P$. The CR system with the random IRS scheme achieves a slightly higher rate than the CR system without IRS by virtue of beamforming with the increased degree-of-freedom of channels. Additionally, the CR system with the proposed IRS outperforms other reference schemes due to the optimal reflection of the IRS. In other words, the proposed IRS dynamically adjusts the reflecting coefficients of IRSs such that the reflected signals propagate toward the SU-RX instead of PU-RX. Furthermore, if more IRSs are deployed, a higher achievable rate can be achieved in the SU-RX.

\begin{figure}[!t]
\centering
    \includegraphics[width=0.99\columnwidth]{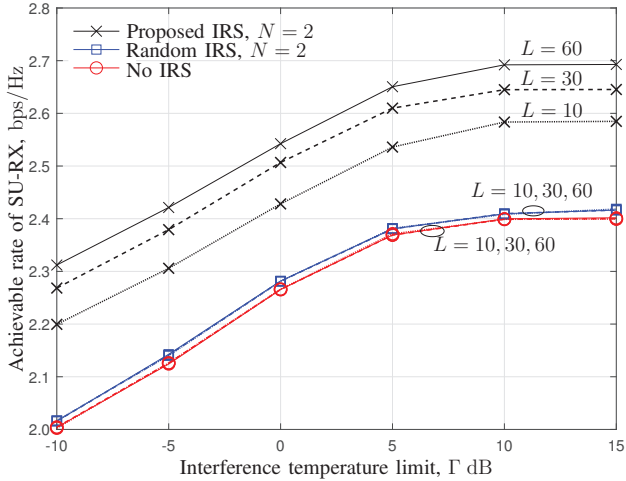}
  \caption{Rate of SU versus different interference temperature limits $\Gamma$ for various numbers of reflecting elements $L\in\{10, 30,60\}$, when $P=10\;{\rm dB}$.}\label{fig_perfect}
\end{figure}

\begin{figure}[!t]
\centering
    \includegraphics[width=0.99\columnwidth]{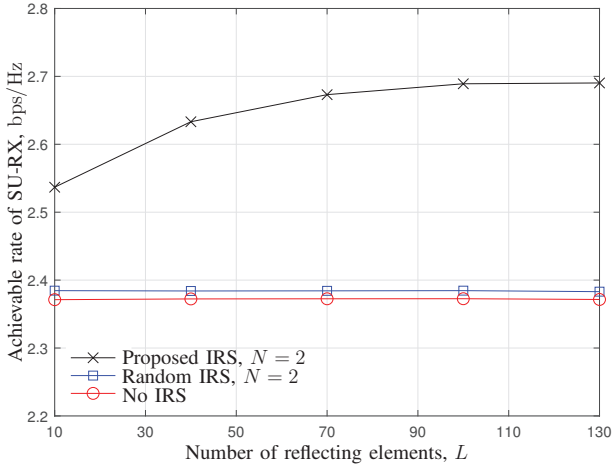}
  \caption{Rate of SU among different passive reflecting elements $L$, when $\Gamma=7\;{\rm dB}$ and $P=10\;{\rm dB}$.}\label{fig_element}
\end{figure}

In Fig. \ref{fig_perfect}, the achievable rates are evaluated over the IT limit $\Gamma$ for varying $L$ when $P=10\;{\rm dB}$. As the interference constraint is more relaxed, i.e., as $\Gamma$ increases, the achievable rate of all schemes increases. Increasing the number of elements of IRS, i.e., $L$, dose not improve the achievable rate of the random-IRS and no-IRS CR systems. In contrast, we observe that the achievable rate of the CR system with the proposed IRSs significantly improves as $L$ increases because the IRSs can be more flexibly designed to focus on an arbitrary direction with stronger reflection when $L$ is large. To further clarify this, in Fig. \ref{fig_element}, the achievable rates are evaluated over $L$ when $\Gamma=7\;{\rm dB}$ and $P=10\;{\rm dB}$. It is evident that the achievable rate of the CR system with the proposed IRS increases with $L$, while that of others is stable.

\subsection{Imperfect CSI Case}
\begin{figure}[!t]
\centering
    \includegraphics[width=0.99\columnwidth]{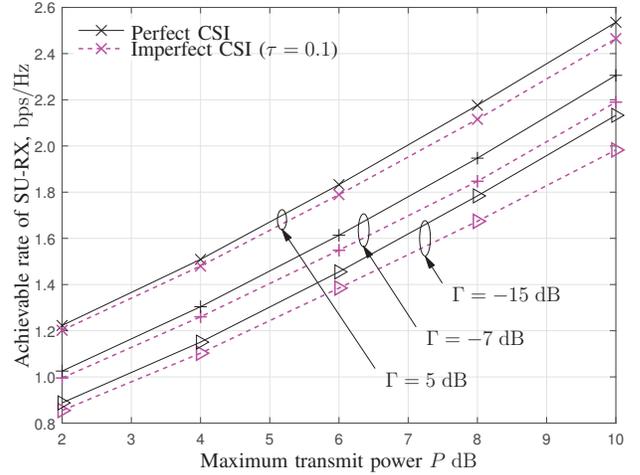}
  \caption{Achievable rate of SU-RX for the IRS-assisted CR system with perfect and imperfect CSI under various IT limits, when $L=10$.}\label{fig_imperfect}
\end{figure}

For the ellipsoidal uncertainty region, we set $\mathbf{C}_{s}=\frac{1}{\xi_s}\mathbf{I}_M$ and $\mathbf{C}_{p,k}=\frac{1}{\xi_{p,k}}\mathbf{I}_M$, i.e., a spherical uncertainty region. Considering the imperfect CSI effects, an {\it uncertainty ratio}, denoted by $\tau$, is defined as follows \cite{li2011optimal}: $\xi_s=\tau\|\mathbf{H}_s\|$ and $\xi_{p,k}=\tau\|\mathbf{H}_{p,k}\|$, where $\tau\in[0,1]$.

In Fig. \ref{fig_imperfect}, the achievable rates of CR systems with perfect CSI and imperfect CSI are plotted as a function of the maximum PU-TX power $P$. Here, different IT limits are considered as $\Gamma\in\{-7,0,7,15\}\;{\rm dB}$. For the imperfect CSI, the uncertainty ratio is set to be $0.01$, i.e., $\tau=0.01$. As expected, the achievable rate is decreased if the CSI is imperfect because the reflected power from the IRS does not accurately focus on the intended SU-RX. One interesting observation is that the performance degradation due to CSI uncertainty becomes trivial as $\Gamma$ increases, i.e., the IT limit becomes relaxed.

\begin{figure}[!t]
\centering
    \includegraphics[width=0.99\columnwidth]{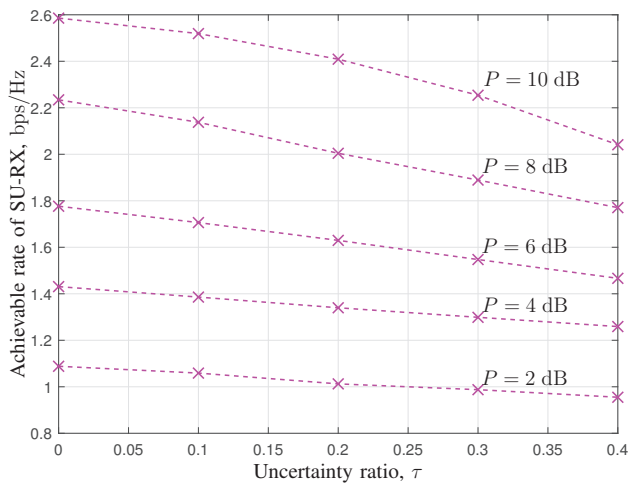}
  \caption{Achievable rate of SU-RX for the IRS-assisted CR system over the CSI uncertainty ratio, when $\Gamma=15\;{\rm dB}$.}\label{fig_noise}
\end{figure}

In Fig. \ref{fig_noise}, the robustness of the proposed IRS against the CSI uncertainty is evaluated. For given $P$, as the CSI uncertainty, i.e., $\tau$, increases, the achievable rate evidently decreases. Specifically, when $P$ is large, the performance degradation is severer. From these results and noting the almost constant improvement in performance, irrespective of $P$, in Fig. \ref{perfect}, we can conclude that the CR system with multiple IRSs would be more beneficial for low-power SU communications, such as Internet of Things (IoT) sensors and devices.

\section{Conclusions}\label{sec:conclusion}


In this study, we have proposed an IRS-assisted CR system to improve the achievable rate of SU without disturbing the existing PU-RXs in the network. To this end, the transmit beamforming vector at the SU-TX and reflecting coefficients at each IRS have been jointly optimized under total transmit power constraint at the SU-TX and IT constraints at the PU-RXs. Besides, both perfect and imperfect CSI cases are considered. We have also verified that the proposed multiple IRSs can significantly improve the achievable rate of SU through rigorous simulation. We would like to mention that the proposed IRS-assisted CR scheme will be of great potential for future machine type communication and worth further study. This scheme provides an overview for combining the low-cost reconfigurable surface with CR network. In our future research, it would be a more interesting problem that IRS can convey its own low rate information by exploiting the spatial dimension.
\bibliographystyle{IEEEtran}
\bibliography{cite_SM}

\end{document}